\documentclass[11pt,a4paper]{article}

\PassOptionsToPackage{dvipsnames}{xcolor}

\usepackage[margin=1in]{geometry}

\usepackage[utf8]{inputenc}
\usepackage[T1]{fontenc}

\usepackage{amsmath,amssymb,amsfonts,mathtools}
\usepackage{bm}

\usepackage[version=4]{mhchem}

\usepackage{algorithm}
\usepackage{algpseudocode}

\usepackage{booktabs}
\usepackage{multirow}
\usepackage{array}
\usepackage{makecell}

\usepackage{graphicx}
\usepackage{subcaption}
\usepackage{float}

\usepackage{tikz}
\usetikzlibrary{arrows.meta, positioning, shapes.geometric, fit, backgrounds, calc}

\definecolor{cBackbone}{HTML}{2E4A6B}
\definecolor{cHead}{HTML}{5A7FA5}
\definecolor{cResolvent}{HTML}{8B5A3C}
\definecolor{cSolver}{HTML}{A8302B}
\definecolor{cElec}{HTML}{4A6FA5}
\definecolor{cOutputBox}{HTML}{3a3a3a}
\definecolor{cEnergy}{HTML}{5F7A3D}

\usepackage[dvipsnames]{xcolor}
\usepackage[colorlinks=true,linkcolor=blue,citecolor=blue,urlcolor=blue]{hyperref}

\usepackage[numbers,sort&compress]{natbib}

\usepackage{enumitem}
\usepackage{siunitx}
\DeclareSIUnit\angstrom{\text{\AA}}
\usepackage{xspace}
\usepackage{cleveref}
\crefname{figure}{Figure}{Figures}
\crefname{equation}{Equation}{Equations}
\crefname{table}{Table}{Tables}
\crefname{section}{Section}{Sections}
\Crefname{figure}{Figure}{Figures}
\Crefname{equation}{Equation}{Equations}
\Crefname{table}{Table}{Tables}
\Crefname{section}{Section}{Sections}

\usepackage{textcomp}

\newcommand{\Etotal}{E_{\mathrm{total}}}
\newcommand{\Eshort}{E_{\mathrm{short}}}
\newcommand{\Eelec}{E_{\mathrm{elec}}}
\newcommand{\Ctilde}{\tilde{C}}
\newcommand{\Qtotal}{Q_{\mathrm{total}}}
\newcommand{\Qstar}{Q^{*}}
\newcommand{\mufrag}{\mu_{\mathrm{frag}}}
\newcommand{\muphys}[1][]{\mu_{\mathrm{phys}\ifx\\#1\\{}\else{,#1}\fi}}
\newcommand{\muglobal}{\mu_{\mathrm{global}}}

\newcommand{\HIPNN}{HIP-NN\xspace}
\newcommand{\hippynn}{\texttt{hippynn}\xspace}
\newcommand{\chieff}{\chi_{\mathrm{eff}}}

\title{Fragment-Constrained Charge Equilibration \\
for Charge-Aware Machine Learning Potentials \\
at Electrochemical Interfaces}

\author{
  Akhil Reddy Peeketi,$^{1,*}$%
  \and Blas P Uberuaga,$^{2}$%
    \and Travis E Jones,$^{1,*}$%
}

\date{
  $^{1}$Theoretical Division, Los Alamos National Laboratory, \\
  Los Alamos, NM 87545, United States of America \\[6pt]
  $^{2}$Materials Science and Technology Division, Los Alamos National Laboratory, Los Alamos, NM 87545, United States of America \\[6pt]
  $^{*}$Corresponding author: \texttt{[peeketiakhilreddy@gmail.com, tejones@lanl.gov]}
}

\begin{document}
\maketitle

\begin{abstract}
\begin{sloppypar}
Predictive simulation of electrochemical interfaces requires atomistic models that capture reactive bond rearrangements, long-range electrostatics, and charge distributions reflecting the electronic distinctness of electrode and electrolyte regions. Existing charge-aware machine-learned interatomic potentials (MLIPs) built on global charge equilibration (QEq) fall short because a single system-wide equilibration condition forces electrode and electrolyte to settle at a common electrochemical potential, leaving no room for the interfacial gradient that the double layer requires and admitting spurious charge transfer between electronically disconnected regions. Per-fragment charge equilibration is the established structural remedy in classical molecular dynamics, but reliance on predefined molecular topology has so far confined it to non-reactive systems. We lift this restriction by making fragment identification itself a differentiable function of atomic geometry, yielding soft fragment-constrained charge equilibration (Soft-FQEq)---a solver layer that restores fragment-resolved charge conservation in reactive MLIPs. The layer consumes four scalar multi-layer perceptron (MLP) readouts from a shared atomic-feature network---per-atom electronegativity, source charge, short-range energy, and a soft bond connectivity---and returns equilibrated charges together with per-fragment chemical potentials. We implement Soft-FQEq as an extension of the \hippynn framework on a \HIPNN feature network and train it on DFT energies, forces, and DDEC6 charges for \ce{IrO2}/\ce{H2O}/\ce{Na+}/\ce{ClO4-} interfaces. The trained model recovers a clear electrode-to-electrolyte gradient in the per-atom electrochemical potential, evaluated as the sum of the learned electronegativity and the Coulomb-summed electrostatic potential at each atom. With the same trained weights but the fragment-constrained solver replaced by global QEq at inference, this interfacial gradient collapses to an essentially uniform profile, directly showing that the gradient cannot be sustained within global QEq while the fragment formulation recovers it. Because the solver couples to the MLIP only through scalar MLP readouts, it provides a route for extending charge-aware potentials to reactive MLIP simulation of electrochemical double layers.
\end{sloppypar}
\end{abstract}

\section{Introduction}
\label{sec:introduction}

Electrochemical interfaces govern the efficiency, selectivity, and stability of technologies ranging from water electrolysis and \ce{CO2} reduction to batteries, supercapacitors, and molten-salt energy systems~\cite{Rossmeisl2007OER,Mom2022IrO2OER,Tian2025CO2RR,Schmuch2018LiBatteries,Salanne2016Supercapacitors,Shen2025SuperSalt}. The electrified solid--liquid interface partitions naturally into an electrode, inner and outer Helmholtz layers of oriented solvent and solvated ions, and a diffuse/bulk electrolyte. The structure of this electric double layer, and its microenvironment of ion correlations, specific adsorption, and interfacial electric fields, determines reactivity and charge storage across these technologies~\cite{Chen2024LocalEnvironment,Gebbie2023EDLActivity,Wu2022EDLReview,Long2025ElectricFields}. Within this structure, the electrochemical potential difference between electrode and solution drives coupled dynamics of electrons, ions, and solvent molecules through bond-breaking and bond-forming events on picosecond-to-nanosecond timescales. Predictive atomistic modeling of such systems therefore requires, simultaneously, reactive chemistry at near-quantum accuracy, environment-dependent partial charges, long-range Coulomb interactions under periodic boundary conditions, and a spatially resolved double layer separating electrode and electrolyte.

Ab initio molecular dynamics (AIMD) based on density functional theory (DFT) can describe reactive interfacial chemistry but is limited by accuracy/cost trade-offs to hundreds of atoms and tens of picoseconds, well below the scales needed to sample realistic double-layer structure and dynamics~\cite{Sundararaman2017GrandCanonical,Hormann2019GrandCanonical,Gross2021GrandCanonical,Sundararaman2022AccurateSim,Gross2022AbInitioMetalWater,Melander2024CIPDFT}. At the opposite end of the cost hierarchy, classical molecular dynamics (MD) with constant-potential electrode methods reaches nanosecond timescales and resolves double-layer structure at aqueous interfaces, but uses static non-reactive force fields for the electrolyte~\cite{Siepmann1995CPM,Jeanmairet2022EDLCSim,Kim2026ConcentratedEDL}. Machine-learning interatomic potentials (MLIPs)~\cite{Bartok2010GAP,Shapeev2016MTP,Smith2017ANI,Zhang2018DeePMD,Drautz2019ACE,Batatia2022MACE,Batzner2022E3,Musaelian2023Allegro} push toward near-DFT accuracy at classical-MD spatial and temporal scales, but standard local MLIPs carry no explicit partial charges or long-range electrostatics. Adding these ingredients defines the class of charge-aware MLIPs~\cite{Ko2021_4GHDNNP_AccChemRes,Anstine2023LongRange,Bergmann2025MLIPDoubleLayer,Jinnouchi2025MLFFElectrochem}, the focus of this work.

Charge equilibration (QEq)~\cite{Rappe1991QEq} solves for self-consistent atomic charges from per-atom electronegativities $\chi_i$ under a global charge-conservation constraint $\sum_i q_i = \Qtotal$; charge-aware MLIPs replace Rappe and Goddard's fixed per-element $\chi_i$ with neural-network predictions of environment-dependent electronegativities~\cite{Ghasemi2015CENT,Ko2021_4GHDNNP}, combining QEq with a range of feature networks and electrostatic solvers~\cite{Ko2021_4GHDNNP,Kocer2025iQEq,Fuchs2025CELLI,Gao2025PQEqFoundation,Weber2026MPNICE,Li2025ShadowMD,Stanton2025ShadowMD,Kaymak2025GPUQEq,Gubler2024PMEQEq,Chalykh2025Electrolytes}. Despite differences in architecture, all such methods share the same stationarity condition,
\begin{equation}
  \chi_i + [\mathbf{A}\mathbf{q}]_i = \muglobal \qquad \text{for every atom } i \text{ in the system,}
  \label{eq:global_qeq}
\end{equation}
where $\mathbf{A}$ is the Coulomb/hardness matrix and $\muglobal$ is a single system-wide Lagrange multiplier enforcing the total-charge constraint $\sum_i q_i = \Qtotal$. \Cref{eq:global_qeq} couples all $N$ atoms (electrode atoms, solvent molecules, and ions alike) to the same scalar $\muglobal$ and therefore cannot support a spatial gradient in $\chi_i + [\mathbf{A}\mathbf{q}]_i$, the quantity we identify in \cref{sec:theory} as the model's local electrochemical potential.

This global-QEq condition cannot describe an electrochemical interface: it forces electrode and electrolyte to equilibrate to a common electrochemical potential $\muglobal$, even though the electrochemical double layer requires a spatial separation in this potential between them. The off-diagonal Coulomb terms between electrode and solution simultaneously transmit the physical electrostatic coupling between regions and open a charge-transfer channel between them; global QEq cannot distinguish these two roles, and charge delocalizes across electronically disconnected regions. This pathology persists under any parameterization of $\bm{\chi}$, $\bm{\eta}$ (per-element hardnesses), or polarization terms. Vondr\'ak, Reuter, and Margraf~\cite{Vondrak2025QEqLimits} demonstrate this directly: spurious long-range charge transfer and overpolarization persist in static electric fields under state-of-the-art neural fits of $\chi$ and $\eta$, confirming the failure as structural rather than fit-limited. Wei et al.~\cite{Weber2026MPNICE} report the same mechanism in MPNICE for Li-ion electrolyte reduction, where global QEq delocalizes charge across molecular fragments.

Prior work has responded to this limitation in two broad ways, neither of which resolves the problem within the QEq energy functional. One class keeps QEq as the charge-determining mechanism but parameterizes it with neural networks~\cite{Ko2021_4GHDNNP,Kocer2025iQEq,Fuchs2025CELLI,Weber2026MPNICE}, sometimes augmenting it with auxiliary degrees of freedom---split-charge resistances, ACKS2 fluctuation variables, or core--shell polarizable charges~\cite{Warren2008Superlinear,Nistor2006SQE,Nistor2009SQE,Verstraelen2009SQEOrganic,Verstraelen2011SQEParams,Verstraelen2013ACKS2,Gao2025PQEqFoundation}---to constrain inter-region charge transfer beyond bare QEq; these additions reduce spurious transfer but cannot eliminate it, since the structural single-$\mu$ pathology of \cref{eq:global_qeq} survives any parameterization. The other class abandons QEq for alternative charge representations---direct charge prediction from neural networks, latent Ewald summation, or total-charge-conditioned potentials~\cite{Nebgen2018TransferableCharges,Anstine2025AIMNet2,King2025LES,Wang2025CPMACE}---each sacrificing a different ingredient: physical atomic charges, an explicit long-range Coulomb decomposition, or a spatially resolved per-atom interfacial potential. A cleaner path within the QEq energy functional is per-fragment equilibration with one Lagrange multiplier per fragment, which yields per-fragment chemical potentials directly. This approach was first deployed in classical fluctuating-charge force fields~\cite{Rick1994TIP4PFQ,ChelliProcacci2002}, where a per-molecule constraint avoids the long-range charge transfer permitted by global equilibration and reproduces liquid water's gas-to-liquid dipole polarization and static dielectric constant, both missed by fixed-charge potentials. But its reliance on hardwiring molecular topology has confined it to non-reactive systems. The obstacle is defining the fragments: any fixed assignment fails the moment a bond breaks.

\begin{figure}[!htb]
     \centering
     \includegraphics[width=0.9\linewidth]{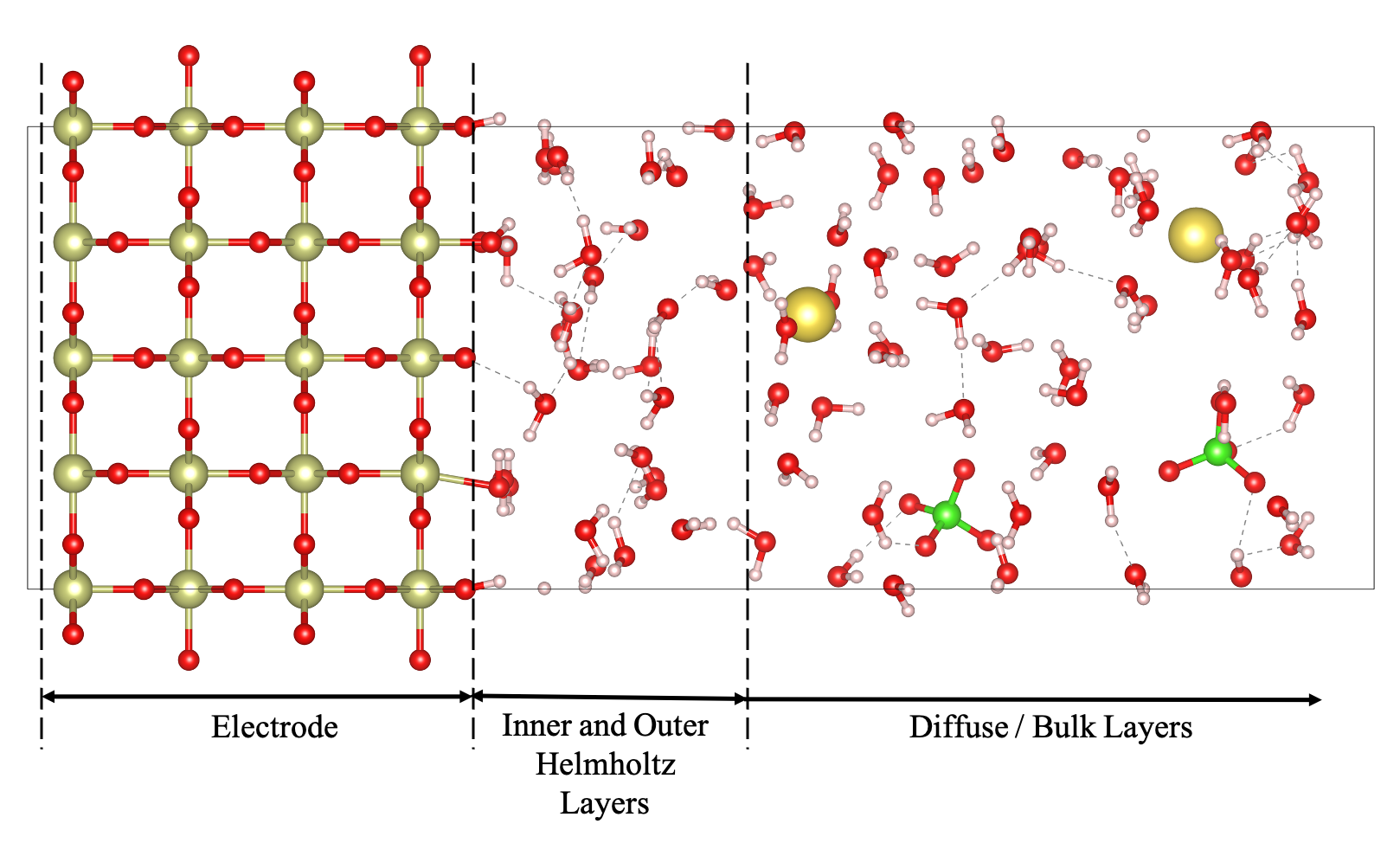}
     \caption{Snapshot of an \ce{IrO2}(110)/aqueous \ce{NaClO4} interface, our prototype training system for Soft-FQEq. The interface partitions into three structurally distinct regions (electrode, inner and outer Helmholtz layers, and diffuse/bulk electrolyte) and, at molecular resolution, into fragments for per-fragment charge equilibration: the connected electrode slab (one fragment), each intact water molecule, and each individual \ce{Na+} or \ce{ClO4-} ion. The electrode (Ir, olive; O, red) is a conductor whose electronic states set the surface's electrochemical potential. The inner and outer Helmholtz layers comprise the first few \si{\angstrom} of solvent: oriented water molecules and, further out, the plane of closest approach of fully solvated ions. The diffuse/bulk region contains the hydrated \ce{Na+} (gold) and \ce{ClO4-} (green, with tetrahedral oxygens) at concentrations approaching those of the bulk electrolyte. Dashed hydrogen bonds are shown in light grey.}
     \label{fig:IrO2_interface}
\end{figure}

We resolve that obstacle by making fragment identification itself a differentiable function of atomic geometry. A small multi-layer perceptron (MLP) readout on per-bond features of a shared atomic-feature network predicts soft bond connectivities (per-bond scalars in $[0,1]$: near 1 for bonded pairs, near 0 for non-bonded pairs, with intermediate values during bond-breaking or bond-forming events); a resolvent operator on the resulting bond graph yields a soft fragment-membership matrix $\Ctilde$ that deforms continuously as atoms move and bonds reorganize. An augmented-Lagrangian Uzawa solver then enforces fragment-resolved charge conservation against this soft membership, producing equilibrated charges and per-fragment chemical potentials $\mufrag$ while remaining differentiable through all iterations. The solver layer couples to the underlying MLIP only through four scalar MLP readouts from the shared atomic-feature network: per-atom electronegativity, source charge, short-range energy, and a soft per-bond connectivity. We refer to this method as Soft-FQEq (soft fragment-constrained charge equilibration) and use this name throughout. We implement Soft-FQEq as an extension of the \hippynn framework~\cite{Lubbers2018HIPNN,Lubbers_hippynn} on a Hierarchically Interacting Particle Neural Network (\HIPNN) feature network and train it on DFT energies, forces, and DDEC6 atomic charges~\cite{Manz2016DDEC6} (chosen for their reproducibility across periodic and molecular systems and their faithful representation of the electrostatic potential) for \ce{IrO2}/\ce{H2O}/\ce{Na+}/\ce{ClO4-} interfaces (\Cref{fig:IrO2_interface}).

The trained model recovers an electrode-to-electrolyte gradient in the per-atom electrochemical potential, evaluated as the sum of the learned electronegativity and the Coulomb-summed electrostatic potential at each atom. Replacing the fragment-constrained solver with global QEq at inference, at the same trained weights, collapses this gradient to an essentially uniform profile, giving a direct architectural demonstration that the interfacial gradient cannot be sustained within global QEq and that fragment constraints are required to express it. The remainder of the paper develops this result in three stages: \cref{sec:theory} presents the soft-fragment formulation, the resolvent construction, the Gaussian-screened Ewald treatment, and the training-time gradient pathways that make the learned heads identifiable, together with a schematic of the full data flow; \cref{sec:methods} describes the DFT reference, the bootstrap training set, the active-learning loop, and the multi-group optimizer; and \cref{sec:results} reports the proof-of-concept validation run, showing the neutral-slab electrochemical-potential profile, the response to systematic electrode charging, and the inference-time ablation.

\section{Theory}
\label{sec:theory}

The total energy takes the form
\begin{equation}
  \Etotal = \Eshort + \Eelec,
  \label{eq:energy_decomposition}
\end{equation}
following the standard short-range + Coulomb split introduced for neural-network potentials in Artrith, Morawietz, and Behler~\cite{Artrith2011NNPCharged} and adopted across subsequent charge-aware MLIP families~\cite{Unke2019PhysNet,Ko2021_4GHDNNP,Anstine2025AIMNet2,Weber2026MPNICE}. The short-range term $\Eshort$ is a per-atom energy predicted by an MLP readout head on the \HIPNN features; following the standard decomposition, we take $\Eshort$ to depend entirely on the local geometric environment and to be independent of the atomic charges, so charge gradients need not be unrolled through the short-range channel. The electrostatic term $\Eelec$ is the fragment-constrained charge-equilibration energy developed in the remainder of this section.

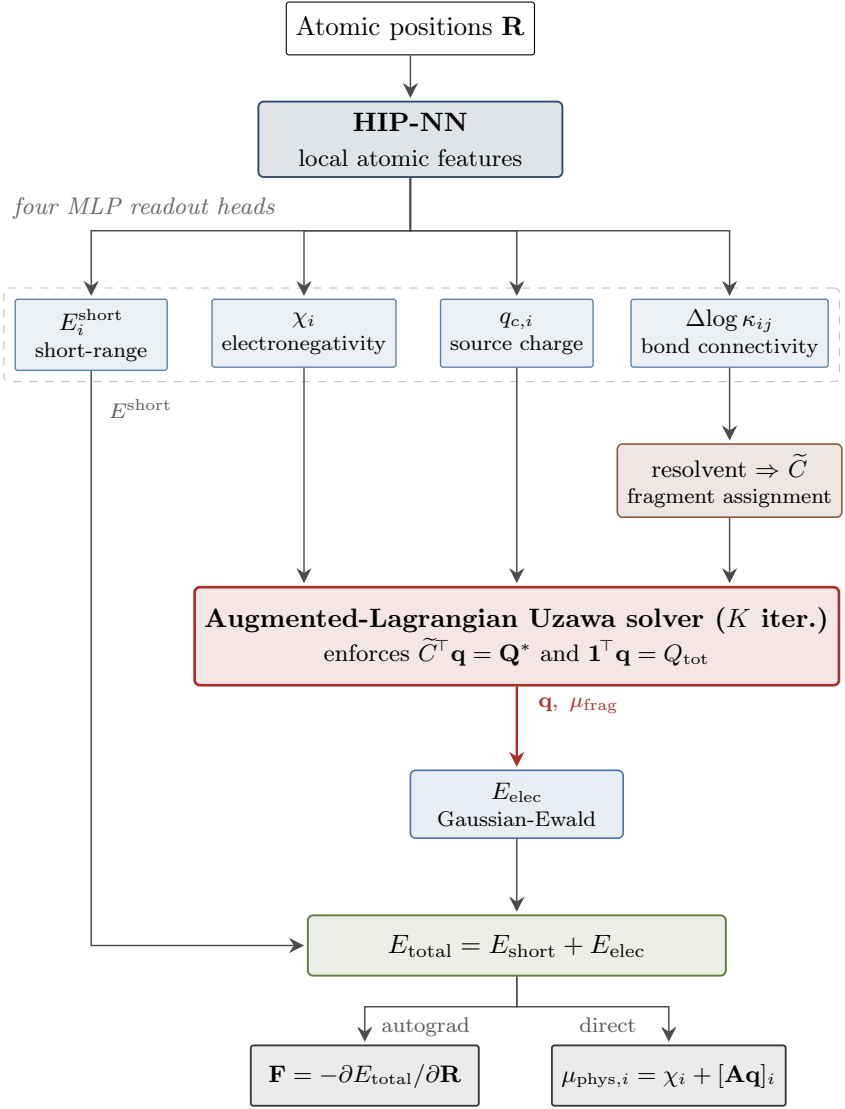
\begin{figure}[!htb]
  \centering
  \begin{tikzpicture}[
    font=\small,
    >={Stealth[length=2mm, width=2mm]},
    io/.style   = {draw=black, fill=white, rounded corners=1pt, minimum width=3cm, minimum height=7mm, align=center, font=\small},
    backbone/.style = {draw=cBackbone, fill=cBackbone!15, line width=0.7pt, rounded corners=2pt, minimum width=4cm, minimum height=10mm, align=center, font=\small\bfseries},
    head/.style  = {draw=cHead, fill=cHead!12, line width=0.5pt, rounded corners=1pt, minimum width=2cm, minimum height=9mm, align=center, font=\footnotesize},
    resolvent/.style = {draw=cResolvent, fill=cResolvent!15, line width=0.6pt, rounded corners=2pt, minimum width=2.8cm, minimum height=9mm, align=center, font=\footnotesize},
    solver/.style = {draw=cSolver, fill=cSolver!12, line width=1pt, rounded corners=2pt, minimum width=6cm, minimum height=13mm, align=center, font=\small\bfseries},
    elec/.style  = {draw=cElec, fill=cElec!12, line width=0.6pt, rounded corners=2pt, minimum width=2.8cm, minimum height=9mm, align=center, font=\footnotesize},
    energy/.style = {draw=cEnergy, fill=cEnergy!12, line width=0.7pt, rounded corners=2pt, minimum width=5.5cm, minimum height=8mm, align=center, font=\small},
    output/.style = {draw=cOutputBox, fill=cOutputBox!10, line width=0.7pt, rounded corners=1pt, minimum width=3cm, minimum height=8mm, align=center, font=\footnotesize},
    flow/.style  = {->, line width=0.55pt, draw=black!70, shorten >=1pt},
    flowthick/.style = {->, line width=0.9pt, draw=cSolver, shorten >=1pt},
  ]

  \node[io] (input) at (0,0) {Atomic positions $\mathbf{R}$};
  \node[backbone, below=6mm of input] (backbone) {HIP-NN \\[1pt] {\normalfont\footnotesize local atomic features}};

  \node[head, below=16mm of backbone.south, xshift=-4.2cm] (hesh)   {$E^{\mathrm{short}}_i$ \\[-1pt] {\scriptsize short-range}};
  \node[head, below=16mm of backbone.south, xshift=-1.4cm] (hchi)   {$\chi_i$ \\[-1pt] {\scriptsize electronegativity}};
  \node[head, below=16mm of backbone.south, xshift=1.4cm]  (hqc)    {$q_{c,i}$ \\[-1pt] {\scriptsize source charge}};
  \node[head, below=16mm of backbone.south, xshift=4.2cm]  (hkappa) {$\Delta\!\log\kappa_{ij}$ \\[-1pt] {\scriptsize bond connectivity}};

  \node[resolvent, below=10mm of hkappa] (resolvent) {resolvent $\Rightarrow$ $\widetilde{C}$ \\[-1pt] {\scriptsize fragment assignment}};

  \node[solver, below=9mm of resolvent, xshift=-2.8cm] (solver) {Augmented-Lagrangian Uzawa solver ($K$ iter.) \\[1pt] {\normalfont\footnotesize enforces $\widetilde{C}^{\!\top}\mathbf{q}=\mathbf{Q}^*$ and $\mathbf{1}^{\!\top}\mathbf{q}=Q_{\mathrm{tot}}$}};

  \node[elec, below=11mm of solver] (eelec) {$E_{\mathrm{elec}}$ \\[-1pt] {\scriptsize Gaussian-Ewald}};
  \node[energy, below=10mm of eelec] (etotal) {$E_{\mathrm{total}} = E_{\mathrm{short}} + E_{\mathrm{elec}}$};

  \node[output, below=9mm of etotal.south, xshift=-2cm] (forces)  {$\mathbf{F} = -\partial E_{\mathrm{total}}/\partial\mathbf{R}$};
  \node[output, below=9mm of etotal.south, xshift=2cm]  (muphys)  {$\mu_{\mathrm{phys},i} = \chi_i + [\mathbf{A}\mathbf{q}]_i$};

  \draw[flow] (input) -- (backbone);
  \foreach \h in {hesh, hchi, hqc, hkappa} {
    \draw[flow] (backbone.south) -- ++(0,-0.7) -| (\h.north);
  }
  \draw[flow] (hkappa) -- (resolvent);
  \draw[flow] (resolvent.south) -- (solver.north -| resolvent.south);
  \draw[flow] (hchi) -- (solver.north -| hchi);
  \draw[flow] (hqc) -- (solver.north -| hqc);
  \draw[flow] (hesh.south) |- (etotal.west);
  \draw[flowthick] (solver.south) -- (eelec.north);
  \draw[flow] (eelec.south) -- (etotal.north);
  \draw[flow] (etotal.south) -- ++(0,-0.4) -| (forces.north);
  \draw[flow] (etotal.south) -- ++(0,-0.4) -| (muphys.north);

  \begin{pgfonlayer}{background}
    \node[draw=black!25, dashed, rounded corners=2pt, fit=(hesh)(hchi)(hqc)(hkappa), inner sep=4pt] (headsgroup) {};
  \end{pgfonlayer}
  \node[font=\footnotesize\itshape, text=black!60, anchor=south west] at ($(headsgroup.north west) + (0.0, 0.78)$) {four MLP readout heads};

  \node[font=\scriptsize, text=cSolver, anchor=west] at ($(solver.south) + (0.15, -0.25)$) {$\mathbf{q},\ \mu_{\mathrm{frag}}$};
  \node[font=\scriptsize, text=black!65, anchor=west] at ($(hesh.south) + (0.1, -0.5)$) {$E^{\mathrm{short}}$};
  \node[font=\scriptsize, text=black!65, anchor=north] at ($(etotal.south) + (-1.2, -0.4)$) {autograd};
  \node[font=\scriptsize, text=black!65, anchor=north] at ($(etotal.south) + (1.2, -0.4)$) {direct};

  \end{tikzpicture}
  \caption{Architecture of fragment-constrained charge equilibration on a \HIPNN feature network. Atomic positions $\mathbf{R}$ feed a shared feature network whose four MLP readout heads produce $E^{\mathrm{short}}_i$, $\chi_i$, $q_{c,i}$, and $\Delta\log\kappa_{ij}$. The bond-connectivity head feeds a differentiable resolvent (\cref{sec:resolvent}) producing the soft fragment-membership matrix $\widetilde{C}$; the augmented-Lagrangian Uzawa solver (\cref{sec:alm}) then consumes $\bm{\chi}$, $\mathbf{q}_c$, and $\widetilde{C}$ and returns equilibrated charges $\mathbf{q}$ and fragment chemical potentials $\mu_{\mathrm{frag}}$. Forces are obtained by autograd through the unrolled solver; the per-atom observable $\mu_{\mathrm{phys},i} = \chi_i + [\mathbf{A}\mathbf{q}]_i$ (\cref{sec:energy_forces}) is evaluated directly from the converged $\bm{\chi}$ and $\mathbf{q}$.}
  \label{fig:architecture}
\end{figure}

A single \HIPNN feature network~\cite{Lubbers2018HIPNN} produces local atomic features that feed four MLP readout heads (Linear $\to$ SiLU $\to$ Linear, in contrast to the two-network scheme of Li et~al.~\cite{Li2025ShadowMD}): the per-atom short-range energy $E^{\mathrm{short}}_i$ (hierarchically initialized), the per-atom electronegativity $\chi_i$ (bounded, centered before the solver), the per-atom source charge $q_{c,i}$ (bounded, zero-initialized for delta-learning), and the per-pair bond-connectivity correction $\Delta\log\kappa_{ij}$ (zero-initialized, on symmetric pair features $\mathbf{f}_i + \mathbf{f}_j$). The latter three enter the fragment-constrained solver; $E^{\mathrm{short}}_i$ sums to the $\Eshort$ term of \cref{eq:energy_decomposition}. Because the solver attaches through these four scalar readouts alone, it is architecture-independent in principle: any MLIP providing per-atom and per-pair scalar features~\cite{Batatia2022MACE,Batzner2022E3,Musaelian2023Allegro,Zhang2018DeePMD,Anstine2025AIMNet2} can be coupled to it without changes, though we have not tested other backbones in this work. \Cref{fig:architecture} shows the full data flow.

The remainder of this section develops the components in detail: the fragment-constrained charge-equilibration energy $\Eelec$ (\cref{sec:alm}); the differentiable fragment-membership construction $\Ctilde$ (\cref{sec:resolvent}); the Gaussian-screened Ewald matrix $\mathbf{A}$ (\cref{sec:ewald}); the auxiliary training-time losses that supervise the upstream heads (\cref{sec:gradient_paths}); and the energy, force, and observable evaluation (\cref{sec:energy_forces}).

\subsection{Fragment-Constrained Charge Equilibration}
\label{sec:alm}

The natural generalization of QEq to an electrochemical interface is to equilibrate charges within each fragment separately~\cite{Rick1994TIP4PFQ}, under a single global charge-conservation constraint. For our \ce{IrO2}/\ce{H2O}/\ce{Na+}/\ce{ClO4-} system, fragments are the electrode slab (a single connected piece), each intact water molecule, and each individual ion (\Cref{fig:IrO2_interface}). Writing $\mathbf{M} \in \{0,1\}^{F \times N}$ for a hard fragment assignment ($F$ fragments, $N$ atoms, one row per fragment), the standard per-fragment QEq problem takes the form
\begin{subequations}\label{eq:hard_qeq}
\begin{align}
  \min_{\mathbf{q}} \quad & \Eelec^{0} = \bm{\chi}^\top \mathbf{q} + \tfrac{1}{2}\,\mathbf{q}^\top \mathbf{A}\mathbf{q}, \label{eq:hard_eelec} \\
  \text{s.t.} \quad & \mathbf{1}^\top \mathbf{q} = \Qtotal, \label{eq:hard_global} \\
  & \mathbf{M}\mathbf{q} = \mathbf{Q}^*_F, \label{eq:hard_frag}
\end{align}
\end{subequations}
with $\mathbf{Q}^*_F \in \mathbb{R}^F$ the vector of per-fragment charge totals and $\mathbf{A}$ the Ewald Coulomb matrix (\cref{sec:ewald}). The stationarity condition introduces a fragment-level Lagrange multiplier, and the bordered Karush--Kuhn--Tucker (KKT) system has dimension $(N+F+1)$: $N$ stationarity equations, $F$ per-fragment conservation equations, and one global conservation equation. Each fragment equilibrates internally to its own chemical potential; inter-fragment charge flow is strictly blocked by the $\mathbf{M}\mathbf{q} = \mathbf{Q}^*_F$ constraint. In the equilibrium limit where fragments are well-defined and their charges known, \cref{eq:hard_qeq} is exactly the problem we want to solve.

This formulation, however, fails in reactive molecular dynamics for two related but distinct reasons. First, $F$ itself is not invariant: when a bond breaks or forms (a water molecule dissociates, a proton transfers to the electrode, two molecules merge), the number of fragments changes discretely, and with it the dimension of $\mathbf{M}$, of $\mathbf{Q}^*_F$, and of the entire bordered KKT system. The problem being solved at one MD step is literally not the same problem as at the next step, and the potential-energy surface (PES) has a jump wherever $F$ changes. Second, even at fixed $F$, $\mathbf{M}$ is a categorical assignment: atoms that cross a fragment boundary switch rows discretely. Both discrete changes occur at the configurations (transition states, bond-breaking events, dissolution) where accuracy matters most for reactive chemistry.

What is needed is a replacement for $\mathbf{M}$ that (i) has dimension independent of the chemistry at any MD step, so the constraint system does not change shape under reactions; (ii) varies smoothly with atomic positions, so atoms can migrate between fragments without PES discontinuities; and (iii) depends differentiably on the parameters that determine fragment identity, so the fragment construction can be learned from data. A matrix of fixed dimension whose entries are $C^\infty$-smooth functions of $\mathbf{R}$ satisfies all three.

We therefore replace the hard $F\times N$ assignment $\mathbf{M}$ by a soft $N\times N$ membership matrix $\Ctilde \in \mathbb{R}^{N\times N}$, with one column per atom and continuous entries in $[0,1]$ (construction in \cref{sec:resolvent}). The constrained QEq problem takes the form
\begin{subequations}\label{eq:constrained_qeq}
\begin{align}
  \min_{\mathbf{q}} \quad & \Eelec^{0} = \bm{\chi}^\top \mathbf{q} + \tfrac{1}{2}\,\mathbf{q}^\top \mathbf{A}\mathbf{q}, \label{eq:eelec_0} \\
  \text{s.t.} \quad & \mathbf{1}^\top \mathbf{q} = \Qtotal, \label{eq:global_constraint} \\
  & \Ctilde^\top \mathbf{q} = \Qstar, \label{eq:frag_constraint}
\end{align}
\end{subequations}
with $\Qstar = \Ctilde^\top \mathbf{q}_c \in \mathbb{R}^N$ giving per-column charge targets from the source charges $\mathbf{q}_c$ (distinct from the solved charges $\mathbf{q}$). The bordered KKT system is now $(2N+1)$-dimensional, constant across all chemistries and reaction stages. Sigmoid sharpening (\cref{sec:resolvent}) makes columns of atoms in the same fragment nearly collinear while columns across fragments have small pairwise overlap, so the $N$ soft constraints have approximate numerical rank equal to the number of distinct fragments $F$ and act collectively as $F$ fragment-level constraints.

In the equilibrium limit (no bonds are breaking or forming, $\Ctilde$ approaches a perfectly sharpened per-fragment indicator, and the constraints $\Ctilde^\top\mathbf{q} = \Qstar$ are exactly satisfied), the soft formulation recovers \cref{eq:hard_qeq}. Soft-FQEq is therefore a continuous extension of standard per-fragment QEq, identical to it when reactions are not occurring and the constraints are honored, and well-defined everywhere else.

With the soft formulation in hand, we turn to the solver. A direct $(2N+1)$ KKT solve would be straightforward, but two features of the problem argue against it. End-to-end training requires gradients from the charge-dependent loss to flow back through the solver into $\bm{\chi}$, $\mathbf{q}_c$, and the bond-connectivity parameters that determine $\Ctilde$; this last requirement becomes a criterion on the $\Ctilde$ construction itself in \cref{sec:resolvent}. A pure quadratic penalty $\tfrac{\beta}{2}\|\Ctilde^\top\mathbf{q} - \Qstar\|^2$ on $\Eelec^{0}$ would make the solve differentiable but would couple constraint enforcement to gradient pathways: large $\beta$ enforces the constraints but dominates the system matrix $(\mathbf{A} + \beta\Ctilde\Ctilde^\top)$ and suppresses the $\bm{\chi}$-gradient, while small $\beta$ preserves gradients at the cost of residual violation. An augmented Lagrangian decouples enforcement from gradient flow: a Lagrange multiplier $\mufrag \in \mathbb{R}^N$ (one component per column constraint) takes over constraint enforcement from $\beta$, growing across outer iterations until the violation vanishes, while a moderate $\beta$ provides curvature for the inner solve and preserves gradients to the upstream heads. A $K$-step Uzawa iteration (alternating an inner charge solve at fixed $\mufrag$ with an outer multiplier update) implements this separation, and can be unrolled for autograd.

The resulting electrostatic energy is
\begin{equation}
  \Eelec = \underbrace{\bm{\chi}^\top\mathbf{q} + \tfrac{1}{2}\,\mathbf{q}^\top\mathbf{A}\mathbf{q}}_{\text{standard QEq}} + \underbrace{\mufrag^\top(\Ctilde^\top\mathbf{q} - \Qstar)}_{\text{Lagrange}} + \underbrace{\tfrac{\beta}{2}\,\|\Ctilde^\top\mathbf{q} - \Qstar\|^2}_{\text{penalty}},
  \label{eq:alm}
\end{equation}
and the Uzawa iteration is given in \cref{alg:uzawa}.

\begin{algorithm}[htbp]
  \caption{Uzawa iteration for fragment-constrained QEq.}
  \label{alg:uzawa}
  \begin{algorithmic}[1]
    \State $\mufrag \gets \mathbf{0}$ \Comment{Cold start}
    \For{$k = 1, \ldots, K$}
      \State $\chieff \gets \bm{\chi} + \Ctilde\,\mufrag - \beta\,\Ctilde\Ctilde^\top \mathbf{q}_c$
      \State $\mathbf{S} \gets \mathbf{A} + \beta\,\Ctilde\Ctilde^\top$
      \State Solve $\begin{bmatrix} \mathbf{S} & \mathbf{1} \\ \mathbf{1}^\top & 0 \end{bmatrix} \begin{bmatrix} \mathbf{q} \\ \tilde{\mu} \end{bmatrix} = \begin{bmatrix} -\chieff \\ \Qtotal \end{bmatrix}$
      \State $\mufrag \gets \mufrag + \alpha\,(\Ctilde^\top \mathbf{q} - \Qstar)$
    \EndFor
    \State \Return $\mathbf{q},\; \tilde{\mu},\; \mufrag$
  \end{algorithmic}
\end{algorithm}

The Lagrange multiplier $\tilde{\mu}$ returned by the bordered solve carries the sign convention of the saddle-point formulation; for consistency with the physical chemical-potential convention adopted in subsequent equations, we define $\muglobal \equiv -\tilde{\mu}$. The converged KKT conditions,
\begin{subequations}\label{eq:kkt}
\begin{align}
  \mathbf{A}\mathbf{q} + \bm{\chi} + \Ctilde\mufrag &= \muglobal\,\mathbf{1}, \label{eq:kkt_stationarity} \\
  \mathbf{1}^\top \mathbf{q} &= \Qtotal, \\
  \Ctilde^\top \mathbf{q} &= \Qstar,
\end{align}
\end{subequations}
yield $\muphys[i] \equiv \chi_i + [\mathbf{A}\mathbf{q}]_i = \muglobal - [\Ctilde\mufrag]_i$. We use $\muphys$ as the model's local electrochemical-potential observable; only region-to-region differences are physically meaningful, because the absolute offset is gauge-dependent. Region-averaged contrasts in $\muphys$ across the electrode--electrolyte boundary are set by $\Ctilde\mufrag$, giving a spatial interfacial gradient consistent with electrochemical double-layer structure. Setting $\Ctilde\mufrag = \mathbf{0}$---equivalently, solving with a global-QEq bordered system that retains only the $\mathbf{1}^\top\mathbf{q} = \Qtotal$ constraint---recovers uniform $\muphys$ everywhere.

Forces on atomic positions are obtained by autograd through all $K$ unrolled Uzawa iterations, taking $-\partial\Etotal/\partial\mathbf{R}$ with $\Etotal = \Eshort + \Eelec$ evaluated at the $K$-step output. At the exact inner minimum of $\Eelec$, variational stationarity (the envelope theorem) would eliminate implicit derivatives through $\mathbf{q}$ because $\partial\Eelec/\partial\mathbf{q} = \mathbf{0}$ there. At finite $K$ this cancellation is incomplete, so the Lagrange-multiplier term of \cref{eq:alm} contributes a non-vanishing constraint force that maintains charge separation between fragments. Omitting $\mufrag$ from the energy while retaining it in the charge solve removes these constraint forces from the gradient, producing large force errors at the electrode--electrolyte boundary.

The penalty strength $\beta$ enters the energy functional and must remain fixed between training and inference. At our chosen $\beta = 50$, the suppression of $\bm{\chi}$-gradients through $(\mathbf{A} + \beta\Ctilde\Ctilde^\top)^{-1}$ is fragment-size-dependent, being empirically much stronger for small molecular fragments than for the large electrode fragment. We use $K = 3$ iterations with step size $\alpha = 0.5$ during training; inference uses larger $K$ (\cref{sec:energy_forces}). The remaining object in the formulation, the soft fragment membership matrix $\Ctilde$, is constructed next.

\subsection{Differentiable Fragment Identification}
\label{sec:resolvent}

Constructing $\Ctilde$ proceeds in two stages: pairwise bond weights from interatomic distances, followed by a fragment-identification pipeline on the resulting bond graph. The bond weights take the form
\begin{align}
  a_\mathrm{base} &= \sigma\!\left(\frac{r_0 - r_{ij}}{\delta}\right) \cdot \sigma\!\left(\frac{r_\mathrm{cut} - r_{ij}}{w_\mathrm{cut}}\right), \label{eq:abase} \\
  W_{ij} &= \sigma\!\bigl(\mathrm{logit}(a_\mathrm{base}) - \Delta\log\kappa_{ij}\bigr), \label{eq:W}
\end{align}
with $\mathrm{logit}(x) = \log[x/(1-x)] = \sigma^{-1}(x)$. Here $r_0$ (target bond length) and $\delta$ (sigmoid width) are fixed per-pair-type parameters calibrated from DFT bond lengths, $r_\mathrm{cut} = r_0 + 3\delta$ is the pair-specific outer cutoff, and $w_\mathrm{cut}$ is a shared width; $\Delta\log\kappa_{ij}$ is the bond-connectivity correction head, with positive values weakening the bond and negative values strengthening it. The construction is symmetric, $W_{ij} = W_{ji}$, inherited from the symmetric pair features $\mathbf{f}_i + \mathbf{f}_j$ that feed the $\kappa$ head. Values of $W_{ij}$ near 1 indicate covalent connectivity, values near 0 indicate no bond, and intermediate values correspond to bond-forming or bond-breaking events. The $\kappa$ head includes a transition-region gate that concentrates corrections near the bond/non-bond decision boundary and an output clamp limiting the bond-midpoint shift; training uses a delayed start that lets $\bm{\chi}$ and $\mathbf{q}_c$ converge first.

From bond weights $\mathbf{W}$, the fragment membership matrix $\Ctilde$ must be constructed to satisfy five criteria simultaneously: (i)~\emph{correctness} (atoms in the same molecule must have identical fragment membership), (ii)~\emph{uniformity} (membership weights within a fragment should be equal, not biased by atom degree or position), (iii)~\emph{differentiability} (fragment assignment must transition smoothly during bond breaking without discrete reassignment), (iv)~\emph{gradient flow} (the $\kappa$ network must receive backpropagation gradient through $\Ctilde$ to be supervised by the charge-dependent loss, per \cref{sec:alm}), and (v)~\emph{global connectivity} (the electrode must be recognized as a single connected fragment; no local bond-distance criterion can determine this).

We evaluated four candidate constructions for $\Ctilde$; each fails at least one criterion. Random walk diffusion $\mathbf{P}^K = (\mathbf{D}^{-1}\mathbf{W})^K$ suffered from severe gradient attenuation: repeated random-walk diffusion damps long-range sensitivity geometrically, and at the $K$ values needed to merge a 150-atom electrode into a single fragment the gradient signal through $\mathbf{P}^K$ to the $\kappa$ network became negligible (criterion iv). Exponential sharpening $\exp(-R^\mathrm{eff}_{ij}/\tau)$ preserves gradients but produces non-uniform $\Ctilde$ columns, penalizing all within-fragment interactions proportionally to graph distance and giving up to 40\% membership variation within the electrode (criterion ii). The heat kernel $\exp(-t\mathbf{L})$, which we expected to sharpen fragment boundaries with increasing $t$, did the opposite: inter-fragment leakage grew with $t$ on connected components with small spectral gaps (criterion i). Hard assignment (binary $\Ctilde$ from distance thresholds) is the simplest option but introduces PES discontinuities at every bond-breaking event (criterion iii), incompatible with reactive MD. The graph Laplacian resolvent, by contrast, admits a construction that satisfies every criterion.

The graph Laplacian resolvent is given by
\begin{equation}
  \mathbf{L} = \mathbf{D} - \mathbf{W}, \qquad
  \mathbf{G} = (\mathbf{L} + \varepsilon\mathbf{I})^{-1}
  \label{eq:resolvent}
\end{equation}
with $\mathbf{D}$ the diagonal degree matrix ($D_{ii} = \sum_j W_{ij}$), $\mathbf{I}$ the identity, and $\varepsilon$ a small regularization. The resolvent is block-diagonal on disconnected fragments (criterion i), attenuates gradients algebraically ($\sim 1/(\lambda + \varepsilon)$) rather than exponentially (criterion iv), is $C^\infty$ smooth in the atomic positions (criterion iii), and encodes the global connectivity of large connected components through the $\sim 1/\varepsilon$ eigenvalue (criterion v). The dense inversion is $O(N^3)$ per forward pass, negligible for our system sizes ($N \lesssim 300$). Uniformity within a fragment (criterion ii) is recovered by sharpening the resolvent through the effective-resistance metric~\cite{Doyle1984RandomWalks}:
\begin{align}
  R_{ij}^\mathrm{eff} &= G_{ii} + G_{jj} - 2G_{ij}, \label{eq:reff} \\
  S^\mathrm{kern}_{ij} &= f_\mathrm{sharp}(R_{ij}^\mathrm{eff}), \label{eq:sigmoid_kern} \\
  \Ctilde &= \mathrm{colnorm}\bigl(\mathbf{G} \odot \mathbf{S}^\mathrm{kern}\bigr), \label{eq:Ctilde}
\end{align}
where $f_\mathrm{sharp}(R) = \sigma\bigl((s_\mathrm{thr} - R)/s_w\bigr)$ is a sigmoid whose threshold $s_\mathrm{thr}$ is chosen above the maximum intra-fragment effective resistance and below the minimum inter-fragment value, $\odot$ denotes elementwise (Hadamard) product, and $\mathrm{colnorm}(\cdot)$ denotes column-normalization to unit sum, $[\mathrm{colnorm}(\mathbf{M})]_{ij} = M_{ij} / \sum_k M_{kj}$. We use $\varepsilon = 0.01$, $s_\mathrm{thr} = 4.0$, and $s_w = 1.0$ throughout. The sharpening preserves $C^\infty$ smoothness, because the sigmoid is flat in both the bonded ($R^\mathrm{eff} \ll s_\mathrm{thr}$) and non-bonded ($R^\mathrm{eff} \gg s_\mathrm{thr}$) regimes, concentrating its variation at the fragment boundary. During a proton transfer, for instance, the hydrogen's column of $\Ctilde$ transitions continuously from donor to acceptor, with no discrete reassignment, fragment counting, or topology update.

\subsection{Gaussian-Screened Ewald Electrostatics}
\label{sec:ewald}

Atoms carry Gaussian charge clouds with per-element widths $\sigma_i$ (set from covalent self-bond lengths, fixed; defined as the Gaussian standard deviation, $\rho_i(\mathbf{r}) \propto \exp[-r^2/(2\sigma_i^2)]$), yielding the screened Coulomb kernel
\begin{equation}
  J_{ij} = k_e \cdot \frac{\mathrm{erf}(r_{ij}/\sqrt{2}\gamma_{ij})}{r_{ij}}, \qquad \gamma_{ij} = \sqrt{\sigma_i^2 + \sigma_j^2},
  \label{eq:gaussian_coulomb}
\end{equation}
with $k_e = \SI{14.4}{\electronvolt\angstrom}$ and a standard Ewald decomposition for periodic boundaries~\cite{Gingrich2010GaussianEwald}. Off-diagonal entries $A_{ij}$ ($i \ne j$) are obtained by Ewald-summing $J_{ij}$ over periodic images, combining a real-space term $k_e[\mathrm{erfc}(\alpha_\mathrm{E} r_{ij}) - \mathrm{erfc}(r_{ij}/\sqrt{2}\gamma_{ij})]/r_{ij}$ with the standard reciprocal-space contribution, where $\alpha_\mathrm{E}$ is the Ewald real/k-space splitting parameter. The diagonal of $\mathbf{A}$ takes the form
\begin{equation}
  A_{ii} = \eta_{Z_i} + \frac{k_e}{\sqrt{\pi}\sigma_i} - \frac{2k_e\alpha_\mathrm{E}}{\sqrt{\pi}},
  \label{eq:diagonal}
\end{equation}
with trainable per-element hardness $\eta_{Z_i}$ initialized from Rappe and Goddard~\cite{Rappe1991QEq}; the second term is the Gaussian self-energy and the third the Ewald self-correction. Gaussian screening is essential to numerical stability: bare $1/r$ Coulomb diverges as $r_{ij} \to 0$ and yields an indefinite $\mathbf{A}$, whereas finite Gaussian widths bound $J_{ij}$ at all distances, with $J_{ij}(r_{ij}\to 0) = k_e\sqrt{2/\pi}/\gamma_{ij}$ set by the combined Gaussian width. With this screening, $\mathbf{A}$ is positive definite with moderate condition number, enabling stable LU-based solves throughout training. We use $\alpha_\mathrm{E} = \SI{0.5}{\per\angstrom}$ with a \SI{7}{\angstrom} real-space cutoff and k-space accuracy target $10^{-5}$.

\subsection{Gradient Pathways and Auxiliary Losses}
\label{sec:gradient_paths}

The primary supervisions are standard: Huber losses on per-atom energies, forces, and solved charges against DFT + DDEC6 references. These alone are insufficient for the fragment-constrained architecture because the Uzawa solver compensates for errors in its upstream heads.

The fragment-constrained solve weakens direct supervision of the upstream $\bm{\chi}$ and $\mathbf{q}_c$ heads. Because the Uzawa iteration can satisfy fragment-level constraints over a broad range of upstream parameters, the solved-charge loss alone does not uniquely determine either electronegativities or source charges: the charge-loss gradient with respect to $\bm{\chi}$ is weak because the solver largely compensates through $\mufrag$, leaving $\bm{\chi}$ without a strong learning signal and at its random initialization. The charges are numerically correct but $\bm{\chi}$ is physically meaningless, the $\Eshort/\Eelec$ decomposition is contaminated, and the $\muphys$ profile becomes an artifact of the random $\bm{\chi}$ initialization rather than a learned observable. The same weak identifiability affects the source charges: the fragment-sum constraint sees only $\Ctilde^\top\mathbf{q}_c$, so any offset $\mathbf{q}_c \to \mathbf{q}_c + \Delta$ with $\Ctilde^\top\Delta = 0$ is invisible to the solver. Auxiliary losses are therefore required to supervise $\bm{\chi}$ and $\mathbf{q}_c$ outside the solver.

A prerequisite for meaningful $\bm{\chi}$ learning is removing the electronegativity gauge freedom. Only differences in $\chi_i$ affect the charges; a uniform shift $\bm{\chi} \to \bm{\chi} + c$ is absorbed by the global Lagrange multiplier $\muglobal$ in the charge conservation constraint, not by $\mufrag$. We center $\bm{\chi}$ to zero mean before entering the solver ($\bm{\chi} \gets \bm{\chi} - \bar{\chi}$), eliminating this degenerate mode and preventing the optimizer from wasting capacity on a physically irrelevant direction.

With the gauge removed, the four auxiliary loss terms take the form
\begin{align}
  L_\chi &= \gamma_\chi\,\mathrm{Var}_f\!\bigl(\bm{\chi} + \mathbf{A}\mathbf{q}_\mathrm{true}\bigr), \label{eq:lchi} \\
  L_{q_c}^{\mathrm{frag}} &= \lambda_q^{\mathrm{f}}\,\bigl\|\Ctilde^\top\mathbf{q}_c - \Ctilde^\top\mathbf{q}_\mathrm{true}\bigr\|^2, \label{eq:lqc_frag} \\
  L_{q_c}^{\mathrm{atom}} &= \lambda_q^{\mathrm{a}}\,\mathrm{Huber}\!\bigl(\mathbf{q}_c,\,\mathbf{q}_\mathrm{true}\bigr), \label{eq:lqc_atom} \\
  L_{\kappa}^{\mathrm{reg}} + L_{\eta}^{\mathrm{reg}} &= \lambda_\kappa\,\|\Delta\log\bm{\kappa}\|^2 + \lambda_\eta\,\|\bm{\eta} - \bm{\eta}_0\|^2. \label{eq:lreg}
\end{align}

$L_\chi$ penalizes within-fragment variation in $\muphys = \bm{\chi} + \mathbf{A}\mathbf{q}_\mathrm{true}$: if $\bm{\chi}$ is correct, atoms in the same fragment should share the same electrochemical potential. With $\Ctilde$ providing soft fragment weights, the fragment-$k$ weighted mean is $\bar{\mu}_k = \sum_i \Ctilde_{ki}\,\muphys[i] / \sum_i \Ctilde_{ki}$, and each atom's membership-weighted prediction is $\hat{\mu}_i = \sum_k \Ctilde_{ik}\,\bar{\mu}_k / \sum_k \Ctilde_{ik}$. $\mathrm{Var}_f(\bm{\mu})$ is the resulting mean squared deviation across atoms, $\mathrm{Var}_f(\bm{\mu}) = (1/N)\sum_i (\muphys[i] - \hat{\mu}_i)^2$. This provides direct gradient to $\bm{\chi}$ without passing through the solver.

The two $\mathbf{q}_c$ losses play complementary roles. $L_{q_c}^{\mathrm{frag}}$ supervises fragment totals $\Ctilde^\top\mathbf{q}_c$ against the DFT reference, fixing what the constrained solve conserves. $L_{q_c}^{\mathrm{atom}}$ supervises the per-atom distribution within each fragment, eliminating the arbitrary offsets that the fragment-sum loss cannot see. Both are required: without the atom-wise term, $\mathbf{q}_c$ is free to develop offsets that cancel on fragment sums but corrupt the intra-fragment structure. This is particularly consequential for the electrode, where the per-atom $\mathbf{q}_c$ anchors the inference-time charge redistribution discussed in \cref{sec:energy_forces}.

The regularizers $L_\kappa^{\mathrm{reg}}$ and $L_\eta^{\mathrm{reg}}$ constrain the scaffolding parameters. $L_\kappa^{\mathrm{reg}}$ keeps the bond-connectivity corrections $\Delta\log\bm{\kappa}$ small, so fragment construction remains dominated by the calibrated distance-sigmoid baseline and the $\kappa$ network acts as a delta correction rather than a free bond classifier. $L_\eta^{\mathrm{reg}}$ anchors per-element hardness near its Rappe--Goddard~\cite{Rappe1991QEq} initialization; unregulated drift reshapes the diagonal of $\mathbf{A}$, trading Coulomb physics for compensator capacity and degrading the condition number of the LU solve.

The task-specific auxiliary losses $L_\chi$, $L_{q_c}^{\mathrm{frag}}$, and $L_{q_c}^{\mathrm{atom}}$ decay as their targets are learned, while the regularizers $L_\kappa^{\mathrm{reg}}$ and $L_\eta^{\mathrm{reg}}$ persist as soft priors on the scaffolding parameters. Together with gauge centering, these terms break the solver's compensatory symmetry and supervise $\bm{\chi}$ and $\mathbf{q}_c$ as independent predictions rather than as solver arguments alone.

The fragment construction parameters $\kappa$ and $\mathbf{q}_c$ also receive gradient through $\Eelec$ itself, via two pathways that hand off during training. Early on, the penalty term of \cref{eq:alm} dominates because the constraint violation is large before $\mathbf{q}_c$ has learned correct targets. As training progresses, the violation shrinks but $|\mufrag|$ grows, and the Lagrange term becomes the dominant gradient source. The shared feature network couples all four readout heads through the force gradient, which propagates $\partial\Eelec/\partial\bm{\chi}$, $\partial\Eelec/\partial\mathbf{q}_c$, and $\partial\Eelec/\partial\kappa$ back into the \HIPNN features.

\subsection{Energy, Forces, and Observables}
\label{sec:energy_forces}

The total energy $\Etotal^{(K)} = \Eshort + \Eelec(\mathbf{q}^{(K)}, \mufrag^{(K)}; \mathbf{R})$ evaluated at the $K$-step Uzawa output yields forces $\mathbf{F} = -\partial\Etotal^{(K)}/\partial\mathbf{R}$ via autograd through all $K$ unrolled Uzawa iterations in float64. Cold-starting $\mufrag = \mathbf{0}$ at every evaluation makes the inner solve a deterministic function of $\mathbf{R}$, preserving a well-defined PES consistent with microcanonical-ensemble integration up to finite-$K$ residuals and integrator error. The penalty strength $\beta = 50$ enters the energy functional and must remain fixed between training and inference; $K$ and $\alpha$ are solver hyperparameters.

The per-atom observable $\muphys[i] = \chi_i + [\mathbf{A}\mathbf{q}]_i$ combines the learned electronegativity $\chi_i$ with the Ewald-summed electrostatic potential at atom $i$ from all charges. It plays the role of a local electrochemical potential on the model's internal reference: a quantity whose spatial variation reports the work required to place a unit charge at each atomic site, combining chemical (electronegativity) and electrostatic contributions as in the thermodynamic definition of electrochemical potential. Only differences in $\muphys$ between spatially separated regions are physically meaningful; the absolute offset is gauge-dependent, set by the centering of $\bm{\chi}$ and the charge-neutral Ewald convention.

The region-averaged difference $\Delta\mu = \mu_{\mathrm{electrode}} - \mu_{\mathrm{bulk}}$ (with $\mu_{\mathrm{electrode}}$ averaged over all electrode atoms and $\mu_{\mathrm{bulk}}$ averaged over $\muphys$ in bulk electrolyte far from the interface) corresponds, on the model's internal scale, to the Galvani (inner) potential difference across the electrode--electrolyte interface. At zero net electrode charge, $\Delta\mu$ is the model's prediction of the intrinsic potential-of-zero-charge offset between the neutral metal--water interface and bulk solvent; its dependence on electrode charging is the capacitive response of the double layer.

Mapping $\Delta\mu$ onto an absolute electrochemical scale (for instance RHE) requires external calibration. A natural route is comparison with reference DFT calculations that treat the bulk solvent as a defined potential reference, such as ESM-RISM~\cite{Nishihara2017ESMRISM}, in which the work function of the slab-electrolyte system is obtained directly from the Fermi energy on the bulk-solvent scale. A single matched-configuration ESM-RISM calculation fixes the additive offset between $\Delta\mu$ and the corresponding work function; within the same electrode--solvent--ion class and comparable concentration regime, this offset is expected to be approximately stable across electrode charge state, coverage, and facet, placing $\Delta\mu$ changes on the RHE scale via the standard conversion. The stability of this offset across configurations---and the known ESM-RISM sensitivity to slab-boundary placement and interfacial water treatment---remains to be validated; the present work reports $\Delta\mu$ on the model's internal scale.

Two aspects of the Uzawa solver differ between training and inference: the iteration count $K$, and the treatment of the autograd through the unrolled iteration. $K$ is 3 during training and 10--20 at inference. Autograd through the solver runs in both cases---for parameter gradients during training, and for forces $-\partial\Etotal^{(K)}/\partial\mathbf{R}$ during MD---but training pays an additional parameter-gradient cost and is sensitive to gradient variance, both of which favor small $K$. Inference is free of these pressures and uses larger $K$ for tighter constraint satisfaction.

The electrode fragment target is the second difference. At training, DDEC6 charge labels are available for every atom, and $L_{q_c}^\mathrm{atom}$ supervises $\mathbf{q}_c$ directly; charge-neutrality of the training cell then implies $\sum_{i\in\mathrm{electrode}} q_{c,i} = -\sum_\mathrm{electrolyte} q_{c,i}$ automatically as the network converges. Training structures need not all contain electrodes: molecular, ionic, and bulk-water configurations are handled by the same loss without special treatment, since the local $L_{q_c}^\mathrm{atom}$ and the fragment-sum $L_{q_c}^\mathrm{frag}$ supervise their $\mathbf{q}_c$ predictions directly from DDEC6. At inference the electrolyte stoichiometry may differ from any training cell, and the local $\mathbf{q}_c$ head (trained from features within a finite cutoff radius) cannot see the global ionic content required to maintain overall neutrality. We therefore set the electrode target by global conservation, which takes the form
\begin{equation}
  Q^*_{\mathrm{electrode}} = \Qtotal - \sum_{f\,\in\,\mathrm{electrolyte}} Q^*_f,
  \label{eq:electrode_by_conservation}
\end{equation}
with $Q^*_f = [\Ctilde^\top\mathbf{q}_c]_f$ for each electrolyte fragment. The Uzawa iteration then uses $\bm{\chi}$ to redistribute $Q^*_{\mathrm{electrode}}$ across the electrode fragment self-consistently, and $L_{q_c}^{\mathrm{atom}}$ ensures this redistribution has a physical per-atom baseline to modulate.

\section{Training Pipeline}
\label{sec:methods}

\subsection{DFT reference calculations}

Spin-polarized DFT calculations were performed with VASP~\cite{Kresse1996VASP} at the PBE-D3(BJ)~\cite{Perdew1996PBE,Grimme2010D3,Grimme2011D3BJ} level with a \SI{450}{\electronvolt} plane-wave cutoff. All calculations are single-point on charge-neutral cells ($\Qtotal = 0$). DDEC6 charges~\cite{Manz2016DDEC6} were computed from the converged all-electron densities using Chargemol~\cite{Limas2018Chargemol}.

\subsection{Bootstrap dataset and active learning}

The bootstrap dataset is constructed to span the full range of local environments the trained model will encounter at an electrochemical interface, from isolated-molecule repulsion and attraction up to extended slab-electrolyte coupling; broader strategies for MLIP data generation are reviewed by Kulichenko et al.~\cite{Kulichenko2024DataGen}. The first of its four subsets, more than 4000 vacuum structures of dimers, trimers, and small molecules drawn from the element set, provides short-range repulsion and attraction baselines. The second, roughly 1000 IrOx bulk structures selected from $\sim 13000$ thermally rattled configurations covering multiple bulk Ir-O phases, is obtained by SOAP-based farthest-point sampling via the DScribe implementation~\cite{Himanen2020DScribe} with a diversity-score cutoff of $\sim 0.005$. The third, approximately 1000 aqueous-electrolyte configurations, is generated by thermally rattling a room-temperature liquid-water snapshot from the revPBE-D3 dataset of Cheng et al.~\cite{Cheng2019IceInWater} with molecule-aware displacements (preserving intramolecular connectivity) and combining the result with a range of cation/anion compositions, then down-selecting the roughly 3000 candidate structures by the same SOAP farthest-point procedure. The fourth, more than 800 unperturbed \ce{IrO2}(110) and \ce{IrO2}(100) slab-electrolyte configurations with varied ion compositions and surface coverages, teaches the model how charge distributes between the electrode, the solvent, and the ionic environment. Together these subsets form a bootstrap set of approximately 6800 structures that provides broad coverage before active learning begins.

Production training extends the bootstrap set through an active learning loop implemented in ALF~\cite{Smith2021ALF}. A structure builder generates randomly perturbed interface configurations; short MD trajectories are run on these under an ensemble of five models; and snapshots are flagged for DFT evaluation where the per-atom energy and force standard deviations across the ensemble exceed configured thresholds. Flagged structures are added to the training set, and the ensemble is retrained. Across successive cycles, two of the five ensemble members are warm-started from the previous cycle and the remaining three are cold-started, preserving the best-performing models from the previous cycle while introducing fresh initializations for uncertainty diversity.

\subsection{Model training}

The training loss combines eight terms. Three are the primary Huber losses on per-atom energies, forces, and solved charges, all computed against the DFT and DDEC6 references. Three are the auxiliary losses $L_\chi$, $L_{q_c}^{\mathrm{frag}}$, and $L_{q_c}^{\mathrm{atom}}$ that supervise $\bm{\chi}$ and $\mathbf{q}_c$ outside the solver (\cref{sec:gradient_paths}). The remaining two, $L_\kappa^{\mathrm{reg}}$ and $L_\eta^{\mathrm{reg}}$, regularize the scaffolding parameters. Optimization uses six parameter groups with separate learning rates and gradient clipping (one each for the \HIPNN feature network, the $\Eshort$ head, the $\bm{\chi}$ head, the $\mathbf{q}_c$ head, the element-wise hardness $\bm{\eta}$, and the $\kappa$-correction head), which accommodates the very different natural learning rates of these heterogeneous parameters. Checkpoints are selected by validation charge loss for the validation run reported in \cref{sec:results}; production active-learning runs use a combined charge and force criterion. The $\mathcal{O}(N^3)$ resolvent and Ewald construction limit practical training to ${\sim}500$ atoms; PME integration~\cite{Kaymak2025GPUQEq,Gubler2024PMEQEq} is planned.

\subsection{Validation run}

The validation reported in \cref{sec:results} is based on a single ensemble training on a curated subset of the bootstrap dataset, spanning all four subsets above. This run is sufficient to establish the architectural behavior: the emergence and collapse of $\Delta\mu$ under fragment and global constraints, the charge-prediction accuracy, and the controls on $\bm{\chi}$ and $\mathbf{q}_c$, but is not production training. Full active-learning production training is in progress.

\section{Validation}
\label{sec:results}

The validation run is a single ensemble training on a curated subset of the bootstrap dataset described in \cref{sec:methods}. The questions we examine here are architectural: whether the model reproduces DDEC6 reference charges, whether the fragment-constrained solve yields a spatial $\muphys$ profile with distinct electrode, interfacial, and bulk-water regions, and whether the electrode chemical potential responds to net-electrode charging as electrochemistry requires. Full active-learning production training is in progress and will be reported separately.

The model reproduces DDEC6 charges with $\sim 0.05\,e$ MAE averaged across elements, with the largest residuals on Cl and Ir reflecting their wider DDEC6 distributions in the training data. The solved charges are the output of the constrained QEq minimization, so this accuracy reflects the joint quality of $\bm{\chi}$, $\mathbf{q}_c$, $\bm{\eta}$, the bond-connectivity network, and the Ewald construction together, not of any single head.

\begin{figure}[!htb]
  \centering
  \includegraphics[width=0.9\linewidth]{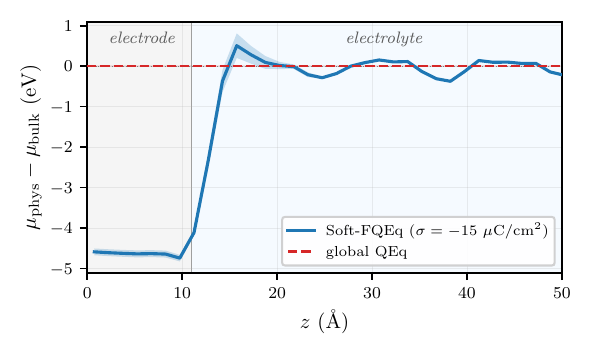}
  \caption{Per-atom $\muphys[i] = \chi_i + [\mathbf{A}\mathbf{q}]_i$ along the interface normal $z$ for a charge-neutral \ce{IrO2}/\ce{H2O}/\ce{NaClO4} slab, referenced to the bulk-water mean ($z \in [30, 60]\,\si{\angstrom}$). The two shaded regions separate the electrode (gray, $z \lesssim \SI{11.5}{\angstrom}$) from the electrolyte (light blue, $z \gtrsim \SI{11.5}{\angstrom}$). Soft-FQEq (blue) shows a flat electrode plateau, a sharp interfacial drop, and a bulk-water plateau near zero; shaded band is the 5-seed ensemble standard deviation. Replacing the Uzawa solver by a global-QEq bordered solve at inference with the same trained weights collapses $\muphys$ to a single system-wide value (red dashed line). Species shown: Ir and electrode-O in the electrode region; clean water-O (not bonded to Cl) in the electrolyte region; H, Na, Cl, and perchlorate oxygens are excluded.}
  \label{fig:mu_profile_neutral}
\end{figure}

\Cref{fig:mu_profile_neutral} shows the per-atom $\muphys[i] = \chi_i + [\mathbf{A}\mathbf{q}]_i$ along the interface normal $z$ for a charge-neutral slab (equal \ce{Na+} and \ce{ClO4-} counter-ions; electrode net charge $Q = 0$). The profile decomposes into three regions. In the electrode ($z \approx 0$--$\SI{12}{\angstrom}$), $\muphys$ is approximately flat---its standard deviation across the Ir sublattice atoms is $\sim\SI{30}{\milli\electronvolt}$, consistent with the structural inequivalence of Ir and O sublattice sites and indicating that within-electrode equilibration is essentially complete. The interfacial region carries a clear gradient in $\muphys$ separating the electrode from bulk water, and in bulk water $\muphys$ reaches a distinct value. The electrode--bulk difference $\Delta\mu = \mu_{\mathrm{electrode}} - \mu_{\mathrm{bulk}}$ is nonzero even at $Q = 0$: this is the potential-of-zero-charge (PZC) offset intrinsic to a neutral metal-water interface, arising from the interfacial dipole layer and water reorientation at the surface. A fragment-constrained solve reproduces this signature because each fragment equilibrates to its own chemical potential; replacing the fragment-constrained solve by global QEq at inference, at the same trained weights, collapses $\muphys$ to a single value uniform to $<\SI{e-6}{\electronvolt}$ across all atoms.

\begin{figure}[!htb]
  \centering
  \includegraphics[width=0.9\linewidth]{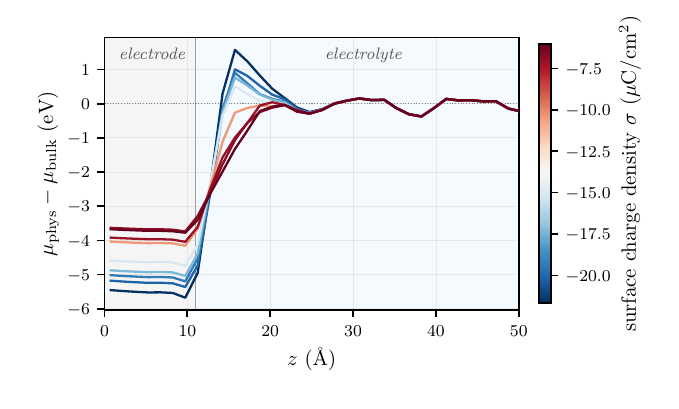}
  \caption{Response of $\muphys(z)$ to electrode charging. Nine ion compositions from 4 \ce{Na+} to 4 \ce{ClO4-} per electrode face (placed in the Stern layer) span a range of model-predicted surface charge densities $\sigma$, with each composition averaged over 5 random seeds. Shaded regions and species inclusion match \cref{fig:mu_profile_neutral}: electrode (gray, $z \lesssim \SI{11.5}{\angstrom}$) and electrolyte (light blue, $z \gtrsim \SI{11.5}{\angstrom}$). The electrode plateau shifts monotonically with $\sigma$: more positive $\sigma$ raises $\muphys$ on the electrode, consistent with the capacitive response expected at an interface. The bulk-water region remains near zero on each configuration's own bulk-water reference. Single-snapshot evaluation per seed, not MD-averaged; curves are 5-seed ensemble means colored by predicted $\sigma$.}
  \label{fig:mu_profile_q_sweep}
\end{figure}

\Cref{fig:mu_profile_q_sweep} shows the response of the $\muphys(z)$ profile to systematic electrode charging. The nine profiles correspond to counter-ion compositions from 4 \ce{Na+} to 4 \ce{ClO4-} per electrode face placed in the Stern layer; each simulation cell remains globally neutral throughout. The model-predicted electrode surface charge density $\sigma$ varies across the sweep, reflecting the partial screening of solution ions by the electrode. The electrode plateau shifts monotonically with $\sigma$: more positive $\sigma$ raises $\muphys$ on the electrode, consistent with the capacitive response expected at an interface. The bulk-water region is less sensitive: the near-bulk plateau close to the interface is approximately $\sigma$-independent, while further into the cell the curves separate by up to a few hundred meV, reflecting a combination of single-snapshot ion-configuration variability and residual long-range polarization from the cell-periodic image. These profiles are computed on single DFT snapshots per seed and ensemble-averaged across 5 seeds; the slope $\partial(\Delta\mu)/\partial\sigma$ is not a measurement of differential capacitance and we show the qualitative response rather than the slope.

Fragment identification is consistent with the underlying chemistry. The model assigns the electrode slab to a single fragment, identifies intact solvent molecules and counter-ion species as separate fragments, and on thermally distorted slabs produces fragment counts consistent with genuine bond dissociation. The kappa-network gate (\cref{sec:resolvent}) is essential: without it, unconstrained $\Delta\log\kappa$ corrections fragment bonded molecules into single-atom pieces, producing fragment totals that have no correspondence to the chemistry and a $\muphys$ profile without physical meaning. All six parameter groups receive nonzero gradient throughout training, and the auxiliary losses $L_\chi$, $L_{q_c}^\mathrm{frag}$, $L_{q_c}^\mathrm{atom}$ decay consistent with their self-annealing construction.

\section{Discussion}
\label{sec:discussion}

The interfacial $\muphys$ gradient emerges in Soft-FQEq because of a specific change to the mathematical structure of QEq, not because the fragment constraints act as an external source of interfacial potential. In global QEq, the stationarity condition $\chi_i + [\mathbf{A}\mathbf{q}]_i = \muglobal$ makes $\muphys[i] \equiv \chi_i + [\mathbf{A}\mathbf{q}]_i$ equal to a single system-wide constant as a structural feature of the equation itself, regardless of how $\bm{\chi}$, $\bm{\eta}$, or the training data are chosen. No quality of the learned parameters can produce a nonzero $\Delta\mu$ against this constraint. In Soft-FQEq, the stationarity condition becomes $\chi_i + [\mathbf{A}\mathbf{q}]_i = \muglobal - [\Ctilde\mufrag]_i$, and $\muphys$ is now permitted to differ across fragments by the amount $[\Ctilde\mufrag]_i$. The Lagrange multipliers $\mufrag$ are the bookkeeping that enforces per-fragment charge conservation and, through the same KKT structure, permits per-fragment chemical potentials to be distinct. We locate the physical content of the double layer---the interfacial polarization, the PZC offset, the response to charging---in the trained $\bm{\chi}$ and the Coulomb field $\mathbf{A}\mathbf{q}$ on the predicted charges; the constraint multipliers are what make this physical content expressible in $\muphys$ rather than being forced to average out to a single global value.

External electrode-charge methods take a different approach to the interfacial problem. In the standard constant-potential method (CPM) as implemented in LAMMPS~\cite{AhrensIwers2022ELECTRODE}, electrode charges are dynamically updated by a capacitance-matrix solve that enforces a prescribed potential on designated electrode atoms, while the electrolyte carries fixed classical force-field charges. This delivers an imposed electrode potential at low computational cost but cannot describe reactive chemistry in the electrolyte: bond breaking, proton-coupled electron transfer, and charge redistribution among solvent and ion species all require electrolyte charges to respond to local environment, which static force-field charges cannot do. Combining CPM with a reactive charge-equilibration method such as the EEM treatment in ReaxFF~\cite{vanDuin2001ReaxFF} makes the electrolyte responsive, but only transfers the problem: the electrolyte still equilibrates to a single chemical potential across water and ions, and the single-$\mu$ pathology analyzed in \cref{sec:alm} persists throughout the solution region. Voltage-equilibration variants~\cite{Onofrio2015EChemDID} impose the electrode potential through a different coupling but share the same electrolyte equilibration pathology. The external electrode constraint only addresses the metal side of the interface; it does not resolve the equilibration pathology within the electrolyte. Soft-FQEq instead treats all subsystems uniformly within one energy functional: electrode, solvent, and each ionic species equilibrate under per-fragment constraints in a single differentiable solve, and forces follow from $\mathbf{F} = -\partial\Etotal/\partial\mathbf{R}$ without external capacitance machinery.

Our ablation isolates architecture from training. The trained weights are held fixed, and at inference the Uzawa solver is replaced by a global-QEq bordered solve. Predicted charges remain close to the DDEC6 reference: the Coulomb operator $\mathbf{A}$ and the trained $\bm{\chi}$ still reconstruct a physically reasonable charge distribution. What collapses is $\muphys$, which falls to a single system-wide value as required by \cref{eq:global_qeq}. Our architectural claim rests on this collapse rather than on any charge-level discrepancy: the interfacial $\muphys$ gradient is ruled out by the stationarity condition of global QEq, independent of whether the trained charges happen to match the reference. Training under global QEq from the start would not rescue the gradient either. The two-parameter $(\chi, \eta)$ flexibility of QEq lets the learned $\bm{\chi}$ absorb the interfacial polarization, but only non-locally: the required correction depends on system-level properties (electrode size, charge state, electrolyte thickness) that lie outside the neural-network cutoff, so any such $\bm{\chi}$ is specific to the training geometry and fails to generalize to different cell sizes or compositions. Vondr\'ak et al.~\cite{Vondrak2025QEqLimits} make the more general point that no $(\bm{\chi}, \bm{\eta})$ parameterization escapes the structural constraint imposed by \cref{eq:global_qeq}. With fragment constraints, the analogous correction lives in $\mufrag$, which is recomputed from the current geometry at every evaluation and adapts to changes in the system without retraining.

We find the $\muphys$ profile robust to random seed only when three components are active together: centering of $\bm{\chi}$ before the solver, the auxiliary loss $L_\chi$ that supervises $\bm{\chi}$ outside the solver, and training on chemically diverse environments that include molecular, bulk-oxide, aqueous-electrolyte, and interfacial configurations. Without $L_\chi$, the solver compensates through $\mufrag$ and $\bm{\chi}$ remains near its random initialization; different seeds produce different $\bm{\chi}$ and therefore different $\muphys$ profiles at matched charge accuracy. Without diverse environments, elements seen in only one chemical context develop systematic electronegativity errors that propagate into the $\muphys$ profile. With all three components active, the electrode-to-bulk gradient is qualitatively consistent across training runs; quantitative agreement across seeds and MD ensembles remains a target for the production run.

Our argument for global-QEq failure is structural: it follows from \cref{eq:global_qeq} alone, not from any feature-specific representation, and applies to every charge-aware MLIP that enforces the single-$\mu$ condition. The fragment-wise remedy is correspondingly a property of the solver layer rather than the feature network: the resolvent, Uzawa iteration, Ewald construction, and augmented Lagrangian consume only predicted per-atom and per-pair scalar features, with no reference to how those features were produced. By the same structural argument, the same four heads and solver layer should apply to other message-passing and equivariant MLIP feature networks (MACE~\cite{Batatia2022MACE}, NequIP~\cite{Batzner2022E3}, Allegro~\cite{Musaelian2023Allegro}, and related architectures) trained end-to-end on DFT and DDEC6 data, with the formulation and training diagnostics---gauge centering, $L_\chi$, $\kappa$-network stabilization, optimizer grouping---transferring while learned parameters do not.

The same logic extends across system classes. Any system whose electronically distinct subsystems are demarcated by absence of covalent bonds inherits the single-$\mu$ pathology of \cref{eq:global_qeq} under global QEq, and the same structural cure under fragment constraints. Concrete candidates include ionic liquids and molten salts (cation and anion as separate fragments, no electrode required to expose the failure mode), aqueous electrolytes without an electrode (each solvated ion as its own fragment), and supported metal clusters on oxide supports (cluster and support as electronically distinct fragments separated at the contact). Each is a domain in which fragment-resolved chemical potentials are physically meaningful but not currently accessible to charge-aware MLIPs.

Our framework provides the ingredients for constant-potential MD---an extractable $\mu_\mathrm{electrode}$, a conservative $\Etotal$, and consistent forces---but does not yet implement it. Constant-charge MD at different ion stoichiometries would yield $\mu_\mathrm{electrode}(Q_\mathrm{electrode})$, the basis for differential-capacitance analysis on dynamical trajectories; on-the-fly GCMC would enable genuine constant-potential sampling. The shadow MD framework~\cite{Li2025ShadowMD,Kaymak2025GPUQEq,Stanton2025ShadowMD}, which propagates charges through an extended Lagrangian, addresses the complementary problem of solver efficiency; composing it with fragment-constrained QEq, propagating both $\mathbf{q}$ and $\mufrag$ as extended variables, is an attractive target.

Several limitations of the present work deserve explicit mention. The $\mathcal{O}(N^3)$ resolvent and Ewald construction currently limit practical training to the hundreds-of-atoms regime; PME integration and block-sparse resolvent approximations for spatially separated fragments are planned. All results reported here are for \ce{IrO2}/\ce{H2O}/\ce{Na+}/\ce{ClO4-} on a \HIPNN backbone; transferability to other electrode materials, solvents, reaction types, or feature-network architectures has not been demonstrated. PES smoothness through reactive events follows from the differentiability of every component of the formulation but has not been empirically validated on bond-breaking trajectories. The $\muphys(z)$ profile has not been compared to planar-averaged DFT Hartree potentials, a straightforward validation that would calibrate the absolute scale and test whether the model captures the interfacial dipole quantitatively.

\section{Conclusion}
\label{sec:conclusion}

The electrochemical double layer at an electrode-electrolyte interface requires architectural support in charge-equilibrated MLIPs. The interfacial $\muphys$ gradient that the double layer requires cannot emerge from global QEq regardless of training quality, parameterization accuracy, or dataset size: the single-$\mu$ condition precludes it mathematically. Fragment-constrained QEq provides this support through the soft constrained problem (\cref{eq:constrained_qeq}), made tractable by a differentiable resolvent that identifies molecular fragments as a continuous function of atomic positions.

The controlled ablation is unambiguous: identical trained weights produce a clear electrode-to-water gradient in $\muphys$ with the fragment-constrained solve, and a single uniform value under global QEq. The architectural fix is structural rather than a training fix: no choice of $(\bm{\chi}, \bm{\eta})$ recovers the interfacial gradient under global QEq, and no tuning of the fragment-constrained solver is needed beyond the formulation itself. The formulation makes spatially resolved electrochemical potentials, fragment-resolved charge states, and an extractable electrode potential accessible in principle to QEq-based MLIPs.

Several forward directions follow directly from the formulation. Production active-learning runs on HPC will test whether the architectural behavior demonstrated on the validation run holds at production accuracy and across a broader set of electrode-coverage and ion-concentration conditions. Constant-charge MD at different ion stoichiometries will map $\mu_\mathrm{electrode}(Q)$ as a dynamical observable rather than a single-snapshot response. On-the-fly GCMC coupled to the extractable $\mu_\mathrm{electrode}$ is the natural route to genuine constant-potential sampling, and multi-electrode cells---two slabs biased at independent potentials within the same simulation---become expressible as two fragment groups with independent Lagrange multipliers. Together these extensions target the full constant-potential MD regime that motivated the fragment-constrained formulation.

\section*{Data Availability}

Training data were generated using VASP at the spin-polarized PBE-D3(BJ) level; DDEC6 charges were computed using Chargemol. The framework extends \hippynn (\url{https://github.com/lanl/hippynn}); software and training configurations will be released upon publication (LA-CC pending).

\section*{Acknowledgments}

The authors acknowledge Laboratory Directed Research and Development (LDRD) Project Number 20240061DR. This research used resources provided by the Los Alamos National Laboratory Institutional Computing Program. Los Alamos National Laboratory is operated by Triad National Security, LLC, for the National Nuclear Security Administration of US Department of Energy (Contract No. 89233218CNA000001). The authors thank B.\ Nebgen for fruitful discussions.

\bibliographystyle{unsrtnat}
\bibliography{references}

\end{document}